\documentclass[a4paper,10pt]{article}

\usepackage[utf8x]{inputenc}
\usepackage{amsmath}
\usepackage{amssymb}
\usepackage{psfrag}
\usepackage[all,cmtip]{xy}
\usepackage{amsmath,amscd, amsthm}
\usepackage{amssymb}
\usepackage{graphicx}
\usepackage{wasysym}
\usepackage{textcomp}
\usepackage{epic,eepic}
\usepackage{graphicx} 
\DeclareGraphicsExtensions{.pdf,.eps,.fig}
\newtheorem{definition}{Definition}
\newtheorem{proposition}{Proposition}

\newtheorem{theorem}{Theorem}
\newtheorem{remark}{Remark}
\newtheorem{example}{Example}

\newcommand{\cat}[1]{\ensuremath{\mathbf{#1}}}
\newcommand{\Cat}[1]{\ensuremath{\mathbf{#1}}}

\newcommand{\after}{\ensuremath{\circ}}
\newcommand{\relto}{\ensuremath{\xymatrix@1@C-2ex{\ar|(.4)@{|}[r] &}}}

\newcommand{\ext}{\ensuremath{^{\mathrm{ext}}}}

\title{Groupoids, Frobenius algebras and Poisson sigma models}
\author{Ivan Contreras}

\begin{document}

\maketitle

\begin{abstract}
This note is devoted to discuss some results proven in \cite{Chris, Relational} and \cite{Thesis} concerning the relation between groupoids and Frobenius algebras specialized in the case of Poisson sigma models with boundary.  We prove a correspondence between groupoids in \textbf{Set} and relative Frobenius algebras in \textbf{Rel},  as well as an adjunction between a special type of semigroupoids and relative H*-algebras.
The connection between groupoids and Frobenius algebras is made explicit by introducing what we called weak monoids and relational symplectic groupoids, in the context of Poisson sigma models with boundary and in particular, describing such structures in the extended symplectic category and the category of Hilbert spaces. 
This is part of a joint work with Alberto Cattaneo and Chris Heunen. 

\end{abstract}

\section{Introduction}
\label{sec:1}
As we know, groupoid structures appear in several scenarios: Lie theory as generalization of Lie groups, in noncommutative 
geometry, foliation theory, Poisson geometry, the study of stacks, among others. On the other hand, Frobenius algebras appear, for example, 
as an equivalent way to understand two dimensional topological quantum field theories (2-TQFT) and it is possible to define them 
in more generality in monoidal dagger categories.\\

In \cite{Chris}, the connection between groupoids and Frobenius algebras is made precise. Namely, there is a way to understand groupoids in the category \textbf{Set} as what we called \emph{Relative Frobenius algebras}, a special type of dagger Frobenius 
algebra in the category \textbf{Rel}, where the objects are sets and the morphisms are relations.\\
In addition, there exists an adjunction between a special type of semigroupoids 
(a more relaxed version of groupoids where the identities or inverses do not necessarily exist) and $H^*-$ algebras, a structure similar to Frobenius algebras 
but without unitality conditions and a more relaxed Frobenius relation.\\

In particular, this correspondence between groupoids and relative Frobenius algebras can be studied in the context of Poisson sigma models (PSM), a particular 2-dimensional topological field theory, where the reduced phase space, for an integrable Poisson manifold $M$ as target space, has the structure of a symplectic groupoid integrating $M$.  In  \cite{Relational}, we study the non reduced phase space of PSM with boundary and we construct what we call a \textit{relational symplectic groupoid}, that is, roughly speaking, a symplectic groupoid up to algebroid homotopy, where the space of morphisms is allowed to be an infinite dimensional weak symplectic manifold and the structure maps of the groupoid are replaced by immersed canonical relations, which are morphisms in the extended symplectic ``category'', denoted by$\mbox{\textbf{Symp}}^{ext}$ 
\footnote{$\textbf{Symp}^{ext}$ is not properly speaking a category, since the composition of canonical relations is not in general a smooth manifold; some transversality conditions are required. For our purposes, the smoothness of the composition of canonical relations will be guaranteed from the defining axioms of the relational symplectic groupoid.}. \\ The study of the non reduced phase space is relevant for the description of general Lagrangian field theories with boundary, following the work of Cattaneo, Mn\"ev and Reshetikhin in \cite{Corfu}. The interesting features of the relational symplectic groupoids could be useful to describe similar constructions in other types of gauge theories.\\

In addition, it turns out that relational symplectic groupoids in the category \textbf{Hilb} of Hilbert spaces correspond to a special type of Frobenius algebras, whereas usual symplectic groupoids in \textbf{Hilb} are in correspondence with relative Frobenius algebras. This would correspond to the quantized version of the relational symplectic groupoid associated to the classical PSM with boundary, assuming that the quantization procedure is functorial.
\section{Groupoids and relative Frobenius algebras}
\label{sec:2}
In this section, we consider a groupoid in \textbf{Set} as a category internal to the category \textbf{Set} of sets as objects and functions as morphisms.
Now, consider the category \textbf{Rel} with sets and relations. In addition, this category carries an involution 
$\dagger: \mbox{\textbf{Rel}}^{op} \to \mbox{\textbf{Rel}}$ given by the transpose of relations;  this is a contravariant involution 
and is the identity on objects, therefore, \textbf{Rel} is a dagger symmetric monoidal category that contains \textbf{Set} as a subcategory. For details on dagger monoidal categories, see e.g. \cite{Abramsky 1, Chris 2}.
In \textbf{Rel} we define what we call \textit{relative Frobenius algebra}, a special dagger Frobenius algebra\footnote{A dagger Frobenius algebra on the category \textbf{Hilb} of finite dimensional Hilbert spaces corresponds to the usual notion of Frobenius algebra.}.
\begin{definition}\emph{
A morphism $m: X\times X \nrightarrow X$ in \textbf{Rel} \footnote{The symbol $\nrightarrow$ denotes that we are considering relations instead of maps as morphisms.} is called a special dagger Frobenius algebra or shortly, relative Frobenius algebra, if it satisfies the following axioms
\begin{itemize}
 \item (F) $(1_X\times m) \circ (m^{\dagger} \times 1_X)= m^{\dagger} \circ m= (m\times 1_X) \circ (1_X \times m^{\dagger}),$
\item (M) $m\circ m^{\dagger}=1_X,$
\item (A) $m\circ (1_X \times m)= m \circ (m\times 1_X),$
\item (U) $\exists u: 1 \nrightarrow X \vert m\circ (u \times 1_X)=1= m\circ(1_X\times u)$.
\end{itemize}
}
\end{definition}
\begin{remark}
\emph{If such $u$ exists, it is unique.} 
\end{remark}

\begin{figure}
\centering
\center{\includegraphics[scale=0.5]{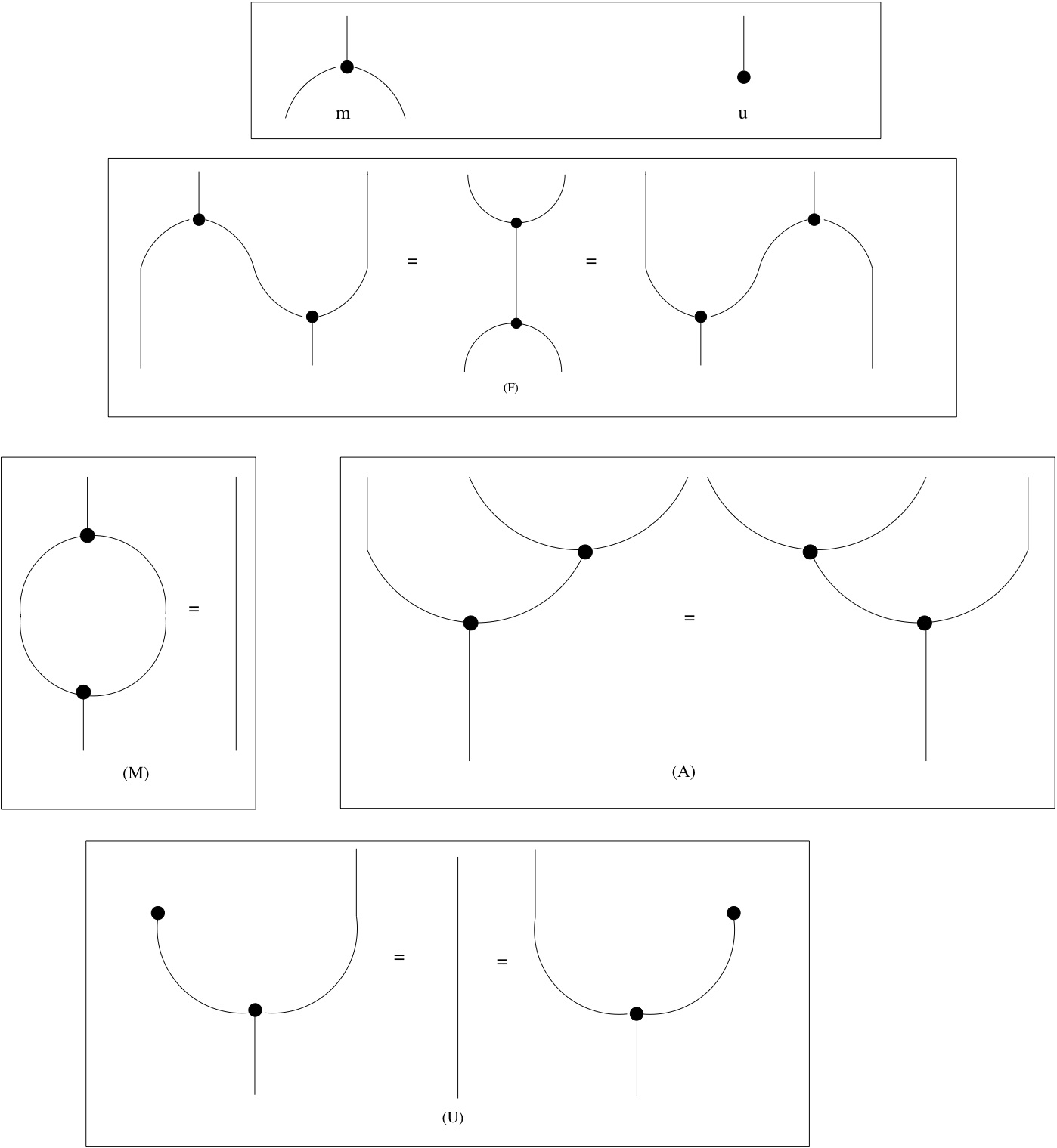}}
\caption{Relative Frobenius algebra: Diagrammatics}
\label{fig:FigureFrobenius}
\end{figure}

\subsection{From relative Frobenius algebras to groupoids}
Here, from a given relative Frobenius algebra we construct a groupoid, but first of all, we give precise meaning of the axioms defined above. We will use the notation $f=hg$ when $((h,g),f)\in m$ and we say that $g$ and $h$ are \emph{composable}. First of all, observe that axiom (M) implies that $m$ is single valued  and 
that 
\[\forall f \in X \, \exists g,\, h \in X \vert f=hg.\]
The axiom (F) means that for all $a,\,b,\,c,\,d\, \in X$
\[ab=cd \Longleftrightarrow \exists \, e \in X \vert b= ed,\, c= ae \Longleftrightarrow \exists \, e \in X \vert d= eb,\, a= ce.\]
The axiom (A) is associativity, i.e. $(fg)h= f(gh)$. For the last axiom, after identifying the morphism $u: 1 \nrightarrow X$ with a subset 
$U \subseteq X$, we get that (U) is equivalent to the following assertions
\begin{center}
\begin{eqnarray*}
\forall f \in X &\exists& \, u \in U \vert fu= f\\
\forall f \in X &\exists& \, u \in U \vert uf= f\\
\forall f \in X  &\forall& \,u \in U \vert f \mbox { and } u \mbox{ are composable} \Longrightarrow fu=f\\
\forall f \in X  &\forall& \,u \in U \vert u \mbox { and } f \mbox{ are composable} \Longrightarrow uf=f.
\end{eqnarray*}
\end{center}

From this data, we are able to give explicitly a groupoid in \textbf{Set}.
\begin{definition}\label{groupoid}
\emph{ Given a relative Frobenius algebra $(X,m)$, we define the following objects and morphisms in \textbf{Rel}:}
\begin{eqnarray*}
G_1&=& X,\\
G_2&=&\{(g,f) \in X^2\vert g \mbox { and } f \mbox{ are composable}\},\\
G_0&=&U,\\
\varepsilon&=& U \times U: G_0 \nrightarrow G_1,\\
s&=&\{(f,x) \in G_1 \times G_0 \vert  f \mbox { and } x \mbox{ are composable}\}: G_1 \nrightarrow G_0\\
t&=&\{(f,y) \in G_1 \times G_0 \vert  y \mbox { and } f \mbox{ are composable} \}: G_1 \nrightarrow G_0\\
\iota&=& \{(g,f) \in G_2 \vert gf \in G_0, \, fg \in G_0 \}: G_1 \nrightarrow G_1.
\end{eqnarray*}
\end{definition}

Using this description of the axioms, it is possible to prove the following
\begin{proposition}
The data 
\\
\xymatrixrowsep{4pc} \xymatrixcolsep{3pc} \xymatrix{
    &\,\,\,\,\;\;\,\,\;\,G_2  \ar[r]^{\,\,\,\,\,\;\;\;\;\;\;\;\;m}  & G_1\ar[r]^{\iota}   &G \ar@/_/[r]_t  \ar@/^/[r]^s & G_0 \ar[l]_{\varepsilon}  & 
    }
\\ 
correspond to a groupoid in \textbf{Set}.
\end{proposition}
\subsection{From groupoids to relative Frobenius algebras}
Here we fix a groupoid 
\\
\xymatrixrowsep{4pc} \xymatrixcolsep{3pc} \xymatrix{
    &\,\,\,\,\;\;\,\,\;\,G_2  \ar[r]^{\,\,\,\,\,\;\;\;\;\;\;\;\;m}  & G_1\ar[r]^{\iota}   &G \ar@/_/[r]_t  \ar@/^/[r]^s & G_0 \ar[l]_{\varepsilon}  & 
    }
\\ 
in \textbf{Set}. 
\begin{definition}\emph{
  For a groupoid $G_1$, define $X=G_1$, and let $m \colon G_1 \times G_1 \nrightarrow G_1$ be
  the graph of the function $m$.}
\end{definition}

We can prove
\begin{proposition}
$(X,m)$ is a relative Frobenius algebra. 
\end{proposition}

Furthermore, under an appropriate choice for morphisms in the corresponding categories, it is possible to prove
\begin{theorem}\label{gpdfrobext}
  There is an isomorphism of categories\emph{ $\mbox{\textbf{Frob(Rel)}}^{ext}
  \cong \mbox{\textbf{Gpd}}^{ext}$}.
\end{theorem}
 The category $\Cat{Gpd}\ext$ has groupoids as objects. Morphisms $\cat{G}
  \to \cat{H}$ are subgroupoids of $\cat{G} \times \cat{H}$.
  The category $\Cat{Frob}(\Cat{Rel})\ext$ has relative Frobenius
  algebras as objects and the choice of the morphisms is natural with respect to the choice of morphisms for $\Cat{Gpd}\ext$, for details see \cite{Chris}
\section{Relative H* -algebras  and semigroupoids}
\begin{definition}
  A \emph{relative H*-algebra is a morphism $m \colon X \times X
  \nrightarrow X$ in $\Cat{Rel}$ satisfying (M), (A), and
  \[
    \tag{H} \begin{array}{c} \text{ there is an involution }* \colon
      \Cat{Rel}(1,X) \to \Cat{Rel}(1,X) \text{ such that } \\
      m \after (1 \times x^*) = (1 \times x) \after m^\dag \text{  and  }
      m \after (x^* \times 1) = (x \times 1) \after m^\dag\\
      \text{ for all } x \colon 1 \nrightarrow X.\end{array}
  \]}
\end{definition}
On the other hand, we have a more relaxed version of groupoids in \textbf{Set}.
A \emph{semigroupoid} consists of a diagram 
\[\xymatrix@C+3ex{
  G_0 
  & G_1 \ar@<-.6ex>|-{s}[l] \ar@<.6ex>|-{t}[l] 
  & G_1 \times_{G_0} G_1 \ar|-{m}[l]
}\]
(in the category $\Cat{Set}$ of sets and functions) such that
\[
  m (m \times 1) = m (1 \times m).
\]
A \emph{pseudoinverse} of $f \in G_1$ is an element $f^* \in G_1$
satisfying ($s(f)=t(f^*)$ and $t(f)=s(f^*)$ and) $f=ff^*f$ and
$f^*=f^*ff^*$. A semigroupoid is \emph{regular} when every $f \in G_1$
has a pseudoinverse. Finally, a semigroupoid is \emph{locally
cancellative} when $fhh^*=gh^*$ implies $fh=g$, and $h^*hf=h^*g$
implies $hf=g$, for any $f,g,h \in G_1$
and any pseudoinverse $h^*$ of $h$.

\subsection{From semigroupoids to relative H*-algebras}

\begin{definition}\emph{
  Given a locally cancellative regular semigroupoid $\cat{G}$, define
  \begin{align*}
    X & = G_1, \\
    m & = \{ (g,f,gf) \mid s(g)=t(f) \} \colon G_1 \times G_1 \nrightarrow G_1, \\
    A^* & = \{a^* \in X \mid a^*aa^*=a^* \mbox{ and } aa^*a=a \mbox{ for all } a \in A \}.
  \end{align*}
  }
\end{definition}

\begin{theorem}
  If $\cat{G}$ is a locally cancellative regular semigroupoid, then
  $m$ is a relative H*-algebra.  
\end{theorem}
\subsection{From H*-algebras to semigroupoids}
\begin{definition}\emph{
  Given a relative H*-algebra $m \colon X \times X \nrightarrow X$, define
  $\cat{G}$ by
  \begin{align*}
    G_0 & = \{ f \in X \mid m(f,f)=f \}, \\
    G_1 & = X, \\
    s & = \{ (f,f^*f) \mid f^* \emph{ is a pseudoinverse of } f \}
    \colon G_1 \nrightarrow G_0 \\
    t & = \{ (f,ff^*) \mid f^* \emph{ is a pseudoinverse of } f \}
    \colon G_1 \nrightarrow G_0. 
  \end{align*}}
\end{definition}
\begin{theorem}\label{locallycancellative}
  If $m$ is a relative H*-algebra, then $\cat{G}$ is a locally
  cancellative regular semigroupoid.
\end{theorem}

  The category $\Cat{LRSgpd}\ext$ has locally cancellative regular
  semigroupoids as objects. Morphisms $\cat{G} \to \cat{H}$ are
  locally cancellative regular subsemigroupoids of $\cat{G} \times \cat{H}$.
In the other hand, the category $\Cat{Hstar}(\Cat{Rel})\ext$ has relative H*-algebras
  as objects and a morphism $(X,m_X) \to (Y,m_Y)$ is a morphism $r \colon
  X \nrightarrow Y$ in $\Cat{Rel}$, natural with respect to the choice of morphisms in $\Cat{LRSgpd}\ext$ \cite{Chris}.
In a similar way as before it can be proven that
\begin{theorem}
  There is an adjunction between $\Cat{LRSgpd}\ext$ and $\Cat{Hstar}(\Cat{Rel})\ext$.
\end{theorem}

\section{Groupoids and Poisson sigma models}
In this section, we describe briefly the construction of groupoids as a way to integrate Poisson manifolds, through the phase space of a 2-dimensional topological field theory, the Poisson sigma model (PSM). This construction was introduced by Cattaneo and Felder in \cite{Ca} and gives explicitly a Lie groupoid  $G \rightrightarrows M$  (for which $G_i$ and $G$ are smooth finite dimensional manifolds  and the structure maps of the groupoid are smooth), if $M$ is an integrable Poisson manifold. In addition, there is a symplectic structure $\omega$ in $G$ that is compatible with the multiplication map $m$; such compatibility turns $G$ into a \textit{symplectic groupoid} integrating the manifold $M$.  More precisely,

\begin{definition}\emph{
A groupoid is called \textit{symplectic} if there is a symplectic structure $\omega$ on $G_1$ such that the graph of $m$ is Lagrangian in $(M, \omega) \times (M , \omega)\times (M, -\omega)$.
}
\end{definition}

The second part of the section is devoted to describe a generalization of such construction, defining what we call a \textit{relational symplectic groupoid}, which lives in the extended symplectic category $\textbf{Symp}^{ext}$, where the objects are (possibly weak) symplectic manifolds and the morphisms are immersed canonical relations.
\footnote{
More precisely, in $\textbf{Symp}^{ext}$, by a morphism between two symplectic manifolds $(M,\omega_M)$ and $(N, \omega_N)$ we mean a pair $(X, p)$ where $X$ is a smooth manifold, $p$ is a smooth map from $X$ to $M\times N$, such that $dp$ is surjective and $T_x(\mathfrak{Im}(p))$ is a Lagrangian subspace of $T(p(x))((M,\omega_M)\times (N, -\omega_N)), \, \forall x\in X$.}.

This construction turns out to be a way to \textit{integrate} any Poisson manifold.
\begin{definition} \emph{A Poisson sigma model (PSM) corresponds to the following data:
\begin{enumerate}
\item A compact surface $\Sigma$, possibly with boundary, called the \textit{source}.
\item A finite dimensional Poisson manifold $(M,\Pi)$, called the \textit{target}. 
\end{enumerate}
}
\end{definition}
The space of fields for this theory is denoted with $\Phi$ and corresponds to the space of vector bundle morphisms between $T\Sigma$ and $T^*M$.
This space can be parametrized by a pair $(X, \eta)$, where $X\in \mathcal C^{k+1}(\Sigma, M)$  and $\eta \in 
\Gamma^k(\Sigma, T^*\Sigma \otimes X^*T^*M),$ and $k \in \{0,\,1,\, \cdots\}$ denotes the regularity type of the map, that we choose to work with.\\
On $\Phi$, the following first order action  is defined:
\[S(X,\eta):= \int_{\Sigma} \langle \eta,\, dX\rangle+  \frac 1 2 \langle \eta, \, (\Pi^{\#}\circ X) \eta \rangle,\]
where,

 \begin{eqnarray}
 \Pi^{\#}: T^*M&\to& TM\\
  \psi &\mapsto& \Pi(\psi, \cdot).
 \end{eqnarray}
 

Here,  $dX$ and $\eta$ are regarded as elements in $\Omega^1(\Sigma, X^*(TM))$, $\Omega^1(\Sigma, X^*(T^*M))$, respectively and  $\langle \,,\, \rangle $ is the pairing between $\Omega^1(\Sigma,  X^*(TM))$ and $\Omega^1(\Sigma,  X^*(T^*M))$ induced by the 
natural pairing between $T_xM$ and $T_x^*M$, for all $x \in M$.

The integrand, called the Lagrangian, will be denoted by $\mathcal{L}$. Associated to this action, the corresponding variational problem $\delta S=0$
induces the following space 
\[\mbox{EL}=\{\mbox{Solutions of the Euler-Lagrange equations}\}\subset \Phi,\]
which is the space of $(X,\eta)$ satisfying the following equations (up to boundary contributions).
\begin{eqnarray}
\frac{\delta \mathcal{L}}{\delta X}&=& dX+ (\Pi^{\#}\circ X)\eta=0\\
\frac{\delta \mathcal{L}}{\delta \eta}&=& d\eta+ \frac 1 2 \langle (\partial \Pi^{\#}\circ X)\eta, \eta \rangle =0.
\end{eqnarray}

Now, if we restrict to the boundary, the general space of boundary fields corresponds to

\[\Phi_{\partial}:=\{\mbox{vector bundle morphisms between  } T (\partial \Sigma) \mbox{  and  } T^*M\}.\]
 
Following the program of classical Lagrangian field theories with boundary,  (\cite{Corfu}), $\Phi_{\partial}$ is endowed with a symplectic form and a surjective submersion $p: \Phi \to \Phi_{\partial}$. We can define 
$$L_{\Sigma}:= p(EL)$$

and also $C_{\Pi}$ as the set of fields in $\Phi_{\partial}$ which  can be completed to a field in  $L_{\Sigma^{'}}$, with $\Sigma^{'}:= \partial \Sigma \times [0, \varepsilon]$, for some $\varepsilon$.  \\
 It turns out that $\Phi_{\partial}$ can be identified with $T^*(PM)$, the cotangent bundle of the path space on $M$ and that
 \[C_{\Pi}:= \{(X,\eta)\vert dX= \pi^{\#}(X)\eta,\, X: \partial \Sigma \to M, \, \eta \in \Gamma (T^*I \otimes X^*(T^*M))\}.\] 
 
It can be proven that  $C_{\Pi}$ is a coisotropic submanifold of finite codimension of $\Phi_{\partial}$ (\cite{Ca}).

\subsection{Geometric interpretation of EL and symplectic reduction}
There is a geometric meaning for the equations of motions of PSM in terms of Lie algebroids that will be useful to understand the reduced phase space in terms of Poisson geometry. In order to do that, we recall some basic notions about Lie algebroids.
\begin{definition}\emph{
A Lie algebroid is a triple $(A, [,]_{A}, \rho)$, where  $\pi: A \to M $ is a vector bundle over $M$,  $[,]_{A}$ is a Lie bracket on $\Gamma(A)$ and $\rho$ (called the anchor map) is a vector bundle morphism from $A$ to $TM$ satisfying the following property\\
\textit{Leibniz property:} \[[X,fY]_{A}=f[X,Y] + \rho_{*}(X)(f)Y, \forall\, X,Y \in \Gamma (A), f \in \mathcal{C}^{\infty}(M).\]
In our case, a basic example of a Lie algebroid is the cotangent bundle of a 
Poisson manifold $T^*M$, where $[,]_{T^*M}$ is the Koszul bracket for 1-forms, that is defined by
$$[df,dg]:=d\{f,g\}, \forall f,g \in \mathcal C ^{\infty}(M),$$
in the case of exact forms and is extended for general 1-forms by Leibniz. 
The anchor map in this example is given by
$\Pi^{\#}: T^*M \to TM$.
}

\begin{definition}\emph{
To define a morphism of Lie algebroids we consider the complex $\Lambda^{\bullet}A^{*}$, where $A^*$ is the dual bundle and a differential $\delta_A$ is defined by 
\begin{eqnarray*}
\delta_A f:&=& \rho^* df,\,\forall f \in \mathcal{C}^{\infty}(M).\\
 \langle \delta_A \alpha, X \wedge Y \rangle&:= &-\langle \alpha,\, [X,Y]_A  \rangle + \langle \delta_A \langle \alpha, X \rangle , Y \rangle\\ &-&\langle \delta_A \langle \alpha, Y \rangle, X \rangle, \, \forall X,Y \in \Gamma(A), \alpha \in \Gamma (A^*),
 \end{eqnarray*}
where $\langle, \rangle$ is the natural pairing between $\Gamma(A)$ and $\Gamma(A^*)$.
A vector bundle morphism $\varphi: A \to B$ is a Lie algebroid morphism if 
\[\delta_A \varphi^*= \varphi^* \delta_B.\]
This condition gives rise to some PDE's that the anchor maps and the structure functions for $\Gamma(A)$ and  $\Gamma(B)$ should satisfy. For the case of PSM, regarding $T^*M$ as a Lie algebroid, we can prove that 
\[C_{\Pi}:=\{\mbox{Lie algebroid morphisms between  } T (\partial \Sigma) \mbox{  and  } T^*M\},\]
where the Lie algebroid structure on the left is given by the Lie bracket of vector fields on $T (\partial \Sigma)$ with identity anchor map.
}
\end{definition}

\end{definition}
Since $C_{\Pi}$ is a coisotropic submanifold, it is possible to perform symplectic reduction, that is, when it is smooth, a symplectic finite dimensional manifold. In the case of $\Sigma$ being a rectangle and with vanishing boundary conditions for $\eta$ (see \cite{Ca}), following the notation in \cite{Cra} and \cite{Severa}, we could also redefine the reduced phase space $\underline{C_{\Pi}}$ as 

\[\underline{C_{\Pi}}:=\left \{ \frac{\mbox{$T^*M$-paths}}{ T^*M \mbox{-homotopy}} \right \}.\]
In the smooth case, it was proven in \cite{Ca} that

\begin{theorem} The following data
\begin{eqnarray*}
G_0&=& M \\
G_1&=& \underline{C_{\Pi}}\\
G_2&=&\{ [X_1, \eta_1], [X_2, \eta_2] \vert X_1(1)=X_2(0)\}
\\
m&:& G_2 \to G:= ([X_1, \eta_1], [X_2, \eta_2]) \mapsto [(X_1* X_2, \eta_1* \eta_2)] \\
\varepsilon&:& G_0 \to G_1:= x\mapsto [X\equiv x, \eta\equiv 0] \\
s&:&G_1 \to G_0:= [X, \eta]\mapsto X(0) \\
t&:&G_1 \to G_0:= [X, \eta]\mapsto X(1) \\
\iota&:& G_1 \to G_1:= [X, \eta] \to [i^* \circ X,i^* \circ \eta]\\
&\mbox{                   }& i:[0,1]\to [0, 1]:= t\to 1-t, 
\end{eqnarray*}
correspond to a symplectic groupoid that integrates the Lie algebroid $T^*M$.  \footnote{here $*$ denotes path concatenation}
\end{theorem}
\subsection{Categorical extensions}
The objective in this section is to introduce several constructions for more general categories (not just $\mbox{\textbf{Symp}}^{ext}$), which resemble the construction of symplectic groupoids and relative Frobenius algebras. More precisely, in the case of Poisson manifolds, the study of the phase space \textit{before reduction} yields to the construction of what we will denote as relational symplectic groupoids. In the sequel we consider a category $\mathcal C$ which admits products and  with a special object $pt$. 

\begin{definition} \emph{A weak monoid in $\mathcal C$ corresponds to the following data:
\begin{enumerate}
\item An object $X$.
\item A morphism $L_1: pt \to X$
\item A morphism $L_3: X\times X \to X,$
\end{enumerate}
satisfying the following axioms
\begin{itemize}
\item (Associativity). $$L_3 \circ (L_3 \times Id)=L_3 \circ (Id \times L_3)$$
\item (Weak unitality). $$L_3\circ(L_1 \times Id)= L_3\circ (Id \times L_1)=:L_2$$
and $L_2\circ L_2=L_2$.
\end{itemize}
We call $L_1$ a weak unit and $L_2$ a projector.
}
\end{definition}
\begin{example}\emph{
Any monoid object in $\mathcal C$ is a weak monoid with $L_1$ being the unit and $L_2$ being the identity morphism. 
}
\end{example}
\begin{example}\emph{
Any relative Frobenius algebra $X$ in \textbf{Rel} is by definition a weak monoid.
}
\end{example}
\begin{example}\emph{
A commutative monoid $(X,m,1)$ equipped with a projector  $p$, that means, $p^2=1$, can be made into a weak monoid. In this case,  $L_3=m$, $L_1=p$ and $L_2: x \mapsto m(p,x)$. Since in general $L_2$ is not the identity morphism, this is not an example of an usual monoid, but for a commutative monoid in \textbf{Set} it can be checked that the quotient $X / L_2$ is a monoid. 
}
\end{example}
\begin{remark}\emph{
The last example does not yield in general to a monoid if we start with a commutative monoid in a category different from \textbf{Set}. For instance, if we take the monoid $\mathbb R, \cdot, 1$ and the projector $p=-1$, the quotient space $\underline X=[0, \infty)$ is a monoid object in \textbf{Set} but it is not an object in \textbf{Man}, the category of smooth manifolds and smooth maps. 
}
\end{remark}
\begin{example}\emph{
It follows from the definition that when $X$ is a vector space, a weak monoid yields into an associative algebra with a preferred central element that induces a projection. This could be called a \textit{prounital associative algebra} (\cite{Thesis}).
}
\end{example}
\begin{definition}
\emph{
Let $\mathcal C$ be a dagger category with products and adjoints.
A weak *-monoid in $\mathcal C$ consists of the following data:
\begin{enumerate}
\item An object $X$
\item A morphism $\psi: X \to X^{\dagger}$
\item A morphism $L_3: X \times X \to X$
\end{enumerate}
such that the following axioms hold
\begin{itemize}
\item (Associativity). $$L_3 \circ (L_3 \times Id)=L_3 \circ (Id \times L_3)$$
\item (Involutivity). $\psi^{\dagger}\psi = Id$
\item  Defining $\psi_R$ the (unique) induced morphism $\psi_R: pt \to X \times X$, then
$$L_1:=L_3\circ \psi_R$$ determines a weak monoid $(X,L_1,L_3)$
\end{itemize}
}
\end{definition}
\begin{example}\emph{
Consider $\mathcal C$ the category $\mbox{\textbf{Vect}} ^{Ext}$ of vector spaces (possibly infinite dimensional) whose morphisms are linear subspaces. The dagger structure is the identity in objects and the relational converse for morphisms. Let $\phi$ be a involutive diffeomorphism of $M$. If $X = \mathcal C^{\infty} (M)$, then $(X, +, \phi^*)$ is a weak *-monoid.
To check this, first observe that 
\begin{eqnarray*}
L_1&=&\{f + \phi^*(f), \, f \in X\}\\
L_2&=&\{(g, g+h+\phi^*h), \, g,h \in X \}\\
L_2 \circ L_2 &=&\{(g, g+h+h^{'}+\phi^*h+\phi^*h^{''}), \, g,h, h^{'} \in X \}.
\end{eqnarray*}
Setting $h^{'}\equiv 0$ we get that $L_2 \subset L_2 \circ L_2$ and by linearity of $\phi$ $L_2 \circ L_2 \subset L_2$. Associativity and unitality follow from the additive structure of $X$.
}
\end{example}
\begin{example}(Deformation quantization).\label{Def}\emph{
Let $\mathcal C=\mbox{\textbf{Vect}} ^{Ext}$ and consider a Poisson manifold $M$. Let $X= \mathcal C ^{\infty}(M, \mathbb C)$ be the algebra of smooth complex valued functions on $M$.  By deformation quantization for Poisson manifolds (see, for example, \cite{Kontsevich}), given a Poisson structure $\Pi$ on $M$, there exists an associative $\mathbb C [\varepsilon]]$- linear product in $X [[ \varepsilon]]$ \footnote{in this case that we are considering complex valued functions we set $\varepsilon= i\hbar / 2$}, denoted by $\star$, such that
\begin{enumerate}
\item $1\star f= f\star 1 =f, \forall f \in X[[ \varepsilon]] $
\item $$f\star g= fg + \varepsilon B_1(f,g)+ \varepsilon^2 B_2(f,g)+\cdots,$$
with $f,\, g \in X \subset X[[\varepsilon]] $ and $B_i$ are bidifferential operators, where 
$$\Pi(df,dg)= \frac{f\star g -g\star f}{\varepsilon}.$$
\end{enumerate}
It can be checked \cite{Thesis} that $(X[[ \varepsilon]], \star, \overline{\cdot})$ is a weak-* monoid, where $\overline{\cdot}$ denotes complex conjugation.
}
\end{example}

\emph{
\begin{definition}\emph{
Let $\mathcal C$ be  a dagger category with products and adjoints.
A cyclic weak *-monoid in $\mathcal C$ consists of the following data:
\begin{enumerate}
\item An object $X$
\item A morphism $\psi: X \to X^{\dagger}$
\item A morphism $L: X \times X \to X^{\dagger}$
\end{enumerate}
such that
\begin{itemize}
\item (Cyclicity). For the associated morphism $L_R: pt \to X^3$
$$L_R= \sigma \circ L_R= \sigma\circ \sigma \circ L_R$$
where 
\begin{eqnarray}
\sigma: X^3 &\to& X^3\\
(a,b,c)&\mapsto& (c,a,b)
\end{eqnarray}
\item If $L_3:= \psi ^{\dagger} \circ L$, then $(X, \psi, L_3)$ is a weak *-monoid.
\end{itemize}
}
\end{definition}
}

\emph{
\begin{example}\emph{(Relational symplectic groupoids). Following \cite{Relational}, we consider $\mathcal C= \mbox{\textbf{Symp}}^{ext}$ and $M$ an arbitrary Poisson manifold.
}
\end{example}
\begin{proposition} The following data
\begin{eqnarray*}
X&:&= T^*(PM)\\
\psi&:& (x, \eta) \mapsto (i^* \circ x,i^* \circ \eta)\\
&\mbox{                   }& i: t \mapsto 1-t\\
L&:&= \{(x_1, \eta_1), (x_1, \eta_1),(x_3, \eta_3) \vert (x_1* x_2, \eta_1* \eta_2) \sim \psi((x_3, \eta_3))\},
\end{eqnarray*}
where $\sim$ denotes the equivalence relation by $T^*M$- homotopy of $T^*M$-paths,
corresponds to a cyclic weak $*$- monoid.
In this case,
\begin{eqnarray*}
L_1&=&\{(x, \eta) \in X \vert (x,\eta)\sim (x\equiv x_0, \eta \equiv 0), x_0 \in M\}\\
L_2&=&\{(x_1, \eta_1), (x_2, \eta_2) \in X \times X \vert (x_1,\eta_1)\sim (x_2, \eta_2)\}.
\end{eqnarray*}
\end{proposition}
}


%
%
%


%

\end{document}